\documentclass[twocolumn,dvipdfmx]{jsiamletters}
\usepackage{color}
\group{
Applied Chaos
}
\affiliation{Department
of Applied Mathematics and Physics, Graduate School of Informatics, Kyoto University}{Yoshida-Honmachi, Sakyo-ku, Kyoto 606-8501, Japan}
\authorinfo{Hirofumi Tsuda}{1}{tsuda.hirofumi.38u@st.kyoto-u.ac.jp}
\authorinfo{Ken Umeno}{1}{umeno.ken.8z@kyoto-u.ac.jp}
\title{Orthogonal Basis Spreading Sequence for Optimal CDMA}
\abstract{Recently, new spreading sequences have been proposed to increase the capacity of users. In particular, the Weyl spreading sequences have the larger capacity of users than the Gold codes. This paper shows that the Weyl spreading sequences appear in a bit recovering model and they are orthogonal basis vectors. This result shows the reason why they have the large capacity and that any spreading sequence is expressed as the sum of the Weyl spreading sequences.}
\usepackage{graphicx}
\usepackage{amsmath}
\begin{document}
\maketitle

\section{Introduction}

The code division multiple access called CDMA is used for 3G mobile communication systems. There are some ways to communicate, for example, frequency-division multiple access and time-division multiple access. In CDMA systems, we use spreading sequences as codes to communicate. CDMA provides a new axis for multiple access, code-division. Recently, OFDM is used as 4G. The number of users has increased and each user has not only one device. In 5G, it is necessary to multiplex to communicate among a number of devices. The frequency spectrum band we can use is limited. Thus, we focus on CDMA again to achieve high capacity.

The current standard spreading sequence is the Gold code \cite{gold}. The new types of spreading sequences have been proposed, for example, the chaos spreading sequence \cite{chaos}, whose SNR equals to the Gold code. The Weyl spreading sequence \cite{weylseq} is based on the Weyl sequence \cite{weyl}, used in a Quasi-Monte Carlo method. Its SNR is larger than the Gold code.

In this paper, we show that the Weyl spreading sequences appear as orthogonal basis vectors in the CDMA model. This is the reason why the Weyl spreading sequences have low interference noise. In Section 2, we show the Weyl spreading sequences and their features. In Section 3, we show the bit recovering model. To recover bits, the value of the cross-correlation is used. This model is a general model since we take into account the value of bits. We divide the cases if consecutive bits are same or different. In both of cases, the Weyl spreading sequences appear as orthogonal basis vectors. In Section 4, we show the new way to decompose spreading sequences. The interference noise is simply expressed with their coefficients of decomposition. 

\section{Weyl Spreading Sequence}
In \cite{weylseq}, the new spreading sequences $\mathbf{w}_k(\sigma)$, the Weyl spreading sequences are proposed by the following formula:
\begin{equation*}
w_{k,n} = \exp\left(2 \pi j \left(n - 1\right) \left(\frac{k}{N} + \sigma\right)\right) \hspace{2mm}(n = 1,2,\ldots,N),
\end{equation*}
where the parameter $\sigma$ satisfies $0 \leq \sigma < 1$, $j$ is an unit imaginary number, $k\hspace{2mm}(1 \leq k \leq K)$ is an integer parameter assigned to each user, and $N$ is the length of the spreading sequences. The complex number $w_{k,n}$ represents the $n$-th element of the user $k$'s spreading sequence $\mathbf{w}_k(\sigma)$. These sequences are based on the Weyl sequences \cite{weyl}. We define the cross-correlation function $C_{k_1,k_2}(l)$ between the different sequences as
\begin{equation}
\begin{split}
C_{k_1,k_2}(l) &= \frac{1}{N}\left\{\sum_{n=1}^{N-l}\overline{w_{k_1,n+l}}w_{k_2,n} \right.\\
 &\left. + b\sum_{n=1}^{l}\overline{w_{k_1,n}}w_{k_2,N-l+n}\right\},
\end{split}
\label{eq:crosscor}
\end{equation}
where $b \in \{-1,1\}$ is a symbol and $\overline{z}$ is the complex conjugate of $z$. Note that (\ref{eq:crosscor}) is a periodic cross-correlation function when $b=1$ and it is an aperiodic cross-correlation function when $b=-1$. These sequences have low cross-correlation functions with different $k$\cite{weylseq}. They satisfy the orthogonal relation as
 \begin{equation*}
 C_{k_1,k_2}(0) = 0
\end{equation*}
for $k_1 \neq k_2$. This result shows that the sequences $\mathbf{w}_k(\sigma)$ are orthogonal to each other.

\begin{figure}[t]
\centering
\includegraphics[scale=0.38]{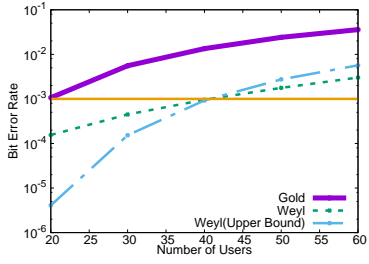}
\caption{Bit Error Rate}
\label{fig:ber}
\end{figure}

It is known in \cite{almost} that the spreading sequences based on the Weyl sequences have the order of the absolute value of cross-correlation $O(1/N)$. The absolute value of cross-correlation with the Gold codes obeys $O(1/\sqrt{N})$. It is desirable that cross-correlation is low since cross-correlation is treated as the interference noise. When the length $N$ gets large, the cross-correlation of the Weyl spreading sequences converges to $0$ faster than one of the Gold codes. In \cite{weylseq}, SNR of the Weyl spreading sequence has been shown as
\begin{equation*}
\mbox{SNR}_{\mbox{Weyl}} \geq \left\{\frac{K-1}{6N} +  \frac{N_0}{2E}\right\}^{-1/2},
\end{equation*}
where $K$ is the number of the users, $E$ is the energy per data bit and $N_0$ is the power of Gaussian noise. In obtaining the upper bound of SNR of the Weyl spreading sequences, it has been assumed that the element $k$ is uniformly distributed in $\{0,1,2,\ldots,N-1\}$. This limitation is efficient when the ratio $K/N$ is close to 1, that is, $K$ is sufficiently large. Fig. \ref{fig:ber} shows numerical BER (Bit Error Rate) result of the Weyl spreading sequence and the Gold code with the length $63$. The Weyl spreading sequences realize much larger multiplexing.

\section{Derivation Orthogonal Basis}
\begin{figure}[htbp] 
   \centering
   \includegraphics[width=2in]{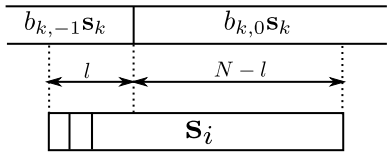} 
   \caption{The model of despreading}
   \label{fig:despreading}
\end{figure}

Let us define the set of all the spreading sequences $S$ with the constant power condition, 
\begin{equation*}
S := \{\mathbf{s} \mid \mathbf{s} \in \mathbb{C}^N,|s_n| = 1\hspace{2mm}(n=1,2,\ldots,N)\}, 
\end{equation*}
where $s_n$ is the $n$-th element of $\mathbf{s}$. The Weyl spreading sequence
\begin{equation*}
\mathbf{w}_k(\sigma) = (w_{k,1}, w_{k,2},\ldots,w_{k,N})^{\mathrm{T}}
\end{equation*}
belongs to $S$. In the above equation, $\mathbf{z}^{\mathrm{T}}$ is the transpose of $\mathbf{z}$.
We assume that the user $k$ has the spreading sequence $\mathbf{s}_k$. We consider a chip-synchronous CDMA model. It is easy to extend this model to the asynchronous model since the interference noise in an asynchronous model is expressed as the sum of the two terms of the interference noise in a chip-synchronous model \cite{pursley}. In CDMA systems, despreading process is necessary to recover bits. Fig. \ref{fig:despreading} shows the model of despreading. The symbols $b_{k,0}, b_{k,-1} \in \{-1,1\}$ are the bits sent by the user $k$. They are spread with the spreading sequence $\mathbf{s}_k$ as in Fig. \ref{fig:despreading}. The value $l$ is the gap length between the beginning of $\mathbf{s}_k$ and the one of $\mathbf{s}_i$. 

Let us define $W_{i,k}(l)$ as the cross-correlation between $\mathbf{s}_i$ and $\mathbf{s}_k$. The cross-correlation $W_{i,k}(l)$ is written as
\begin{equation}
\begin{split}
W_{i,k}(l) =& T_c \exp(j \phi_k)\left\{b_{k,-1}\sum_{n=1}^{l}\overline{s_{i,n}}s_{k,N-l+n}\right. \\
&\left.+ b_{k,0}\sum_{n=1}^{N-l} \overline{s_{i,l+n}}s_{k,n}\right\} ,
\end{split}
\label{eq:cor}
\end{equation}
where $\phi_k \in [0,2\pi)$ is the phase of user $k$'s carrier signal and $T_c$ is the width of each chip. For simplicity, we set $\phi_k=0$ and $T_c=1$. Then, \eqref{eq:cor} is rewritten as the following quadratic form:
\begin{equation}
W_{i,k}(l) = \mathbf{s}^*_i B^{(l)}_{b_{k,-1},b_{k,0}} \mathbf{s}_k,
\label{eq:cor2}
\end{equation}
where
\begin{equation}
B^{(l)}_{b_{k,-1},b_{k,0}} = \left( \begin{array}{c c}
O & b_{k,-1}E_{l} \\
b_{k,0}E_{N-l} & O
\end{array}
\right),
\label{eq:matrix}
\end{equation}
$E_{l}$ is the $l$-dimensional unit matrix, and $\mathbf{z}^*$ is the conjugate transpose of $\mathbf{z}$. Note that $B^{(l)}_{b_{k,-1},b_{k,0}}$ is a normal matrix. This matrix is diagonalizable by the unitary matrices. It is clear that
\begin{equation*}
\begin{split}
B^{(l)}_{1,1} =&  -B^{(l)}_{-1,-1},\\
B^{(l)}_{-1,1} =&  -B^{(l)}_{1,-1}.
\end{split}
\end{equation*}
The matrices $B^{(l)}_{1,1}$ and $B^{(l)}_{-1,-1}$ represent that the same consecutive bits are sent. The matrices $B^{(l)}_{-1,1}$ and $B^{(l)}_{1,-1}$ represent that the different consecutive bits are sent. It is sufficient to consider only $B^{(l)}_{1,1}$ and $B^{(l)}_{-1,1}$.

\subsection*{(A) $B^{(l)}_{1,1}$}
Let us define $\lambda^{(l)}_m$ and $\mathbf{v}^{(l)}_m$ as the $m$-th eigenvalue and the corresponding eigenvector of $B^{(l)}_{1,1}$. They satisfy the relation that
\begin{equation*}
B^{(l)}_{1,1}\mathbf{v}^{(l)}_m = \lambda^{(l)}_m \mathbf{v}_m^{(l)}.
\end{equation*}
From \eqref{eq:matrix}, a vector $\mathbf{z} \in \mathbb{C}^{N}$ is transformed by $B^{(l)}_{1,1}$ as
\begin{equation*}
\left( \begin{array}{c c}
O & E_{l} \\
E_{N-l} & O
\end{array}
\right) \left( \begin{array}{c}
z_1\\
\vdots\\
z_{N-l}\\
z_{N-l+1}\\
\vdots\\
z_N
\end{array}
\right) = \left( \begin{array}{c}
z_{N-l+1}\\
\vdots\\
z_{N}\\
z_{1}\\
\vdots\\
z_{N-l}
\end{array}
\right).
\end{equation*}
Note that $B^{(l)}_{1,1}$ is a permutation matrix. About $B^{(l)}_{1,1}$, the following equation is satisfied for an integer $1\leq l \leq N$:
\begin{equation*}
B^{(l)}_{1,1} = B^{(1)}_{1,1}B^{(l-1)}_{1,1}.
\end{equation*}
Thus, from the above equation, we obtain the relation that
\begin{equation*}
\begin{split}
B^{(l)}_{1,1}\mathbf{v}^{(1)}_m &= B^{(1)}_{1,1}B^{(l-1)}_{1,1}\mathbf{v}^{(1)}_m\\
&=\left(B^{(1)}_{1,1}\right)^l \mathbf{v}^{(1)}_m\\
&=\left(\lambda^{(1)}_m\right)^{l}\mathbf{v}^{(1)}_m.
\end{split}
\end{equation*}
From this relation, the matrices $B^{(l)}_{1,1}\hspace{2mm}(1 \leq l \leq N)$ have the same eigenvector $\mathbf{v}_m$ whose eigenvalue is expressed as the products of $\lambda^{(1)}_m$, that is,
\begin{equation*}
\lambda^{(l)}_m = \left( \lambda^{(1)}_m \right)^l , \mathbf{v}^{(l)}_m = \mathbf{v}_m, 
\end{equation*}
for $1 \leq l \leq N$.
The eigenvalue of $B_{1,1}^{(1)}$, $\lambda^{(1)}_m$ is expressed as
\begin{equation*}
\lambda^{(1)}_m = \exp\left(-2 \pi j \frac{m}{N}\right) \hspace{2mm}(1 \leq m \leq N).
\end{equation*}
Therefore, the eigenvalue $\lambda^{(l)}_m$ and the eigenvector $\mathbf{v}_m$ are expressed as
 \begin{equation}
 \begin{split}
\lambda^{(l)}_m  &= \exp\left(-2 \pi j \frac{ml}{N}\right),\\
 \mathbf{v}_m &= \frac{1}{\sqrt{N}}\left( \begin{array}{c}
 1\\
 \exp(2 \pi j \frac{m}{N})\\
  \exp(2 \pi j \frac{2m}{N})\\
  \vdots\\
   \exp(2 \pi j \frac{(N-1)m}{N})\\
   \end{array}
   \right) = \frac{1}{\sqrt{N}}\mathbf{w}_m(0),
   \label{eq:basis1}
 \end{split}
\end{equation}
where $1 \leq m \leq N$. In the above equation, the Weyl spreading sequence $\mathbf{w}_m(0)$ appears as the eigenvector. From the above result, we can decompose the matrix $B^{(l)}_{1,1}$ and rewrite \eqref{eq:cor2} as
 \begin{equation}
 W_{i,k}(l) = \mathbf{s}^*_i  V \Lambda^{(l)} V^*\mathbf{s}_k,
 \label{eq:mat1}
\end{equation}
where 
 \begin{equation*}
 \begin{split}
V &= \left( \begin{array}{c c c c}
\mathbf{v}_1 & \mathbf{v}_2& \cdots & \mathbf{v}_N
\end{array} \right),\\
\Lambda^{(l)} &= \left( \begin{array}{c c c c}
\lambda^{(l)}_1 & 0 & \cdots & 0 \\
0 & \lambda^{(l)}_2 & \cdots & 0 \\
\vdots & \vdots & \ddots & \vdots \\
0 & 0 & \cdots & \lambda^{(l)}_N
\end{array} \right).
\end{split}
\end{equation*}
From \eqref{eq:basis1} and \eqref{eq:mat1}, $|W_{i,k}(l)|$ gets low if we choose $\mathbf{w}_k(\sigma)$ as $\mathbf{s}_k$ and $i \neq k$. In particular, $W_{i,k}(l) = 0$ if we choose $\mathbf{w}_k(0)$ as $\mathbf{s}_k$.


\subsection*{(B) $B^{(l)}_{-1,1}$}
Similar to (A), let us define $\hat{\lambda}^{(l)}_m$ and $\hat{\mathbf{v}}^{(l)}_m$ as the $m$-th eigenvalue and the corresponding eigenvector of $B^{(l)}_{-1,1}$. They satisfy the relation that
\begin{equation*}
B^{(l)}_{-1,1}\hat{\mathbf{v}}^{(l)}_m = \hat{\lambda}^{(l)}_m \hat{\mathbf{v}}^{(l)}_m.
\end{equation*}
From \eqref{eq:matrix}, a vector $\mathbf{z} \in \mathbb{C}^{N}$ is transformed by $B^{(l)}_{-1,1}$ as
\begin{equation*}
\left( \begin{array}{c c}
O & -E_{l} \\
E_{N-l} & O
\end{array}
\right) \left( \begin{array}{c}
z_1\\
\vdots\\
z_{N-l}\\
z_{N-l+1}\\
\vdots\\
z_N
\end{array}
\right) = \left( \begin{array}{c}
-z_{N-l+1}\\
\vdots\\
-z_{N}\\
z_{1}\\
\vdots\\
z_{N-l}
\end{array}
\right).
\end{equation*}
About $B^{(l)}_{-1,1}$, the following equation is satisfied for an integer $1\leq l \leq N$:
\begin{equation*}
B^{(l)}_{-1,1} = B^{(1)}_{-1,1}B^{(l-1)}_{-1,1}.
\end{equation*}
From the above equation, we obtain the relation that
\begin{equation*}
\begin{split}
B^{(l)}_{-1,1}\hat{\mathbf{v}}^{(1)}_m &= B^{(1)}_{-1,1}B^{(l-1)}_{-1,1}\hat{\mathbf{v}}^{(1)}_m\\
&=\left(B^{(1)}_{-1,1}\right)^l \hat{\mathbf{v}}^{(1)}_m\\
&=\left(\hat{\lambda}^{(1)}_m\right)^{l}\hat{\mathbf{v}}^{(1)}_m.
\end{split}
\end{equation*}
The matrices $B^{(l)}_{-1,1}\hspace{2mm}(1 \leq l \leq N)$ have the same eigenvector $\hat{\mathbf{v}}^{(1)}_m$ whose eigenvalue is expressed as the products of $\hat{\lambda}^{(1)}_m$. The vector $\hat{\mathbf{v}}^{(l)}_m$ is rewritten as $\hat{\mathbf{v}}_m$ since it is independent of $l$. The eigenvalue of $B^{(1)}_{-1,1}$, $\hat{\lambda}^{(1)}_m$ is expressed as
\begin{equation*}
\hat{\lambda}^{(1)}_m = \exp\left(-2 \pi j \left(\frac{m}{N} + \frac{1}{2N}\right)\right) \hspace{2mm}(1 \leq m \leq N).
\end{equation*}
Therefore, the eigenvalue $\hat{\lambda}^{(l)}_m$ and the eigenvector $\hat{\mathbf{v}}_m$ are expressed as
 \begin{equation}
 \begin{split}
 \hat{\lambda}^{(l)}_m  &= \exp\left(-2 \pi j l\left(\frac{m}{N} + \frac{1}{2N}\right)\right) \hspace{2mm},\\
 \hat{\mathbf{v}}_m &= \frac{1}{\sqrt{N}}\left( \begin{array}{c}
 1\\
 \exp(2 \pi j (\frac{m}{N} + \frac{1}{2N}))\\
  \exp(2 \pi j 2(\frac{m}{N} + \frac{1}{2N}))\\
  \vdots\\
   \exp(2 \pi j (N-1)(\frac{m}{N} + \frac{1}{2N}))\\
   \end{array}
   \right)\\
    &= \frac{1}{\sqrt{N}}\mathbf{w}_{m}\left(\frac{1}{2N}\right),
   \label{eq:basis2}
   \end{split}
\end{equation}
where $1 \leq m \leq N$. In the above equation, the Weyl spreading sequence $\mathbf{w}_{m}\left(1/(2N)\right)$ appears as the eigenvector. From the above result, we can decompose the matrix $B^{(l)}_{-1,1}$ and rewrite \eqref{eq:cor2} as
 \begin{equation}
 W_{i,k}(l) = \mathbf{s}^*_i  \hat{V} \hat{\Lambda}^{(l)} \hat{V}^*\mathbf{s}_k,
 \label{eq:mat2}
\end{equation}
where 
 \begin{equation*}
 \begin{split}
\hat{V} &= ( \begin{array}{c c c c}
\hat{\mathbf{v}}_1 & \hat{\mathbf{v}}_2& \cdots & \hat{\mathbf{v}}_N
\end{array} ),\\
\hat{\Lambda}^{(l)} &= \left( \begin{array}{c c c c}
\hat{\lambda}^{(l)}_1 & 0 & \cdots & 0 \\
0 & \hat{\lambda}^{(l)}_2 & \cdots & 0 \\
\vdots & \vdots & \ddots & \vdots \\
0 & 0 & \cdots & \hat{\lambda}^{(l)}_N
\end{array} \right).
\end{split}
\end{equation*}
From \eqref{eq:basis2} and \eqref{eq:mat2}, $|W_{i,k}(l)|$ gets low if we choose $\mathbf{w}_k(\sigma)$ as $\mathbf{s}_k$ and $i \neq k$. In particular, $W_{i,k}(l) = 0$ if we choose $\mathbf{w}_k\left(1/(2N)\right)$ as $\mathbf{s}_k$. 

\section{Expression of Spreading Sequence}
We have obtained the orthogonal basis vectors $(1/\sqrt{N})\mathbf{w}_m(0)$ and $(1/\sqrt{N})\mathbf{w}_m(1/2N)$. Thus, any spreading sequence is expressed as the weighted sum of them. Let us define a spreading sequence $\mathbf{s} \in S$. The sequence $\mathbf{s}$ is expressed as
 \begin{equation*}
 \begin{split}
\mathbf{s} = \frac{1}{\sqrt{N}}\sum_{m=1}^N \alpha_m\mathbf{w}_m(0) = \frac{1}{\sqrt{N}}\sum_{m=1}^N \beta_m\mathbf{w}_m\left(\frac{1}{2N}\right),
\end{split}
\end{equation*}
where $\alpha_m$ and $\beta_m \in \mathbb{C}$ are the coefficients. They satisfy
 \begin{equation*}
 \begin{split}
 \alpha_m = \frac{1}{\sqrt{N}}\left\langle \mathbf{w}_m\left(0\right),\mathbf{s}\right\rangle, \beta_m = \frac{1}{\sqrt{N}}\left\langle\mathbf{w}_m\left(\frac{1}{2N}\right),\mathbf{s}\right\rangle,
 \end{split}
\end{equation*}
where
 \begin{equation*}
 \left\langle\mathbf{s}_i,\mathbf{s}_k\right\rangle = \sum^N_{n=1}\overline{s_{i,n}}s_{k,n}.
\end{equation*}
 Let us define the matrices $\Phi$ and $\hat{\Phi}$ as
 \begin{equation*}
 \begin{split}
 \Phi &= \frac{1}{N}\left( \begin{array}{c c c c}
 \phi_{1,1} &  \phi_{1,2} & \cdots &  \phi_{1,N}\\
  \phi_{2,1} &  \phi_{2,2} & \cdots &  \phi_{2,N}\\
  \vdots & \vdots & \ddots & \vdots\\
   \phi_{N,1} &  \phi_{N,2} & \cdots &  \phi_{N,N}\\
 \end{array} \right), \\
  \hat{\Phi} &= \frac{1}{N}\left( \begin{array}{c c c c}
 \hat{\phi}_{1,1} &  \hat{\phi}_{1,2} & \cdots &  \hat{\phi}_{1,N}\\
  \hat{\phi}_{2,1} &  \hat{\phi}_{2,2} & \cdots &  \hat{\phi}_{2,N}\\
  \vdots & \vdots & \ddots & \vdots\\
   \hat{\phi}_{N,1} &  \hat{\phi}_{N,2} & \cdots &  \hat{\phi}_{N,N}\\
 \end{array} \right),
  \end{split}
 \end{equation*}
 where
 \begin{equation*}
 \begin{split}
 \phi_{p,q} =& \left\langle\mathbf{w}_p(0),\mathbf{w}_q\left(\frac{1}{2N}\right)\right\rangle\\
 =&\frac{2}{1-\exp(2 \pi j (\frac{q-p}{N} + \frac{1}{2N}))},\\
 \hat{\phi}_{p,q} =& \left\langle\mathbf{w}_p\left(\frac{1}{2N}\right),\mathbf{w}_q\left( 0 \right)\right\rangle\\
 =&\frac{2}{1-\exp(2 \pi j (\frac{q-p}{N} - \frac{1}{2N}))}.
  \end{split}
 \end{equation*}
Between $\alpha_m$ and $\beta_m$, there is the relationships
 \begin{equation*}
 \left( \begin{array}{c}
 \alpha_1\\
  \alpha_2\\
  \vdots\\
   \alpha_N
   \end{array} \right) =
 \Phi \left( \begin{array}{c}
 \beta_1\\
  \beta_2\\
  \vdots\\
   \beta_N
 \end{array} \right), \hspace{2mm} \left( \begin{array}{c}
 \beta_1\\
  \beta_2\\
  \vdots\\
   \beta_N
   \end{array} \right) =
 \hat{\Phi} \left( \begin{array}{c}
 \alpha_1\\
  \alpha_2\\
  \vdots\\
   \alpha_N
 \end{array} \right). \nonumber
\end{equation*}
With the above expression, we can easily calculate the function $W_{i,k}(l)$. The coefficients $\alpha^{(k)}_m$ and $\beta^{(k)}_m$ are obtained from $\mathbf{s}_k$. When the user $k$ sends the same bits, for example $1$ and $1$,
 \begin{equation*}
 \begin{split}
 W_{i,k}(l) &= \mathbf{s}^*_i B^{(l)}_{1,1} \mathbf{s}_k\\
 &= \mathbf{s}^*_i V \Lambda^{(l)} V^* \mathbf{s}_k\\
 &=\sum_{m=1}^N \lambda^{(l)}_m \overline{\alpha^{(i)}_m}\alpha^{(k)}_m.
  \end{split}
  \end{equation*}
  When the user $k$ sends the different bits, for example $-1$ and $1$,
 \begin{equation*}
 \begin{split}
 W_{i,k}(l) &= \mathbf{s}^*_i B^{(l)}_{-1,1} \mathbf{s}_k\\
 &= \mathbf{s}^*_i \hat{V} \hat{\Lambda}^{(l)} \hat{V}^* \mathbf{s}_k\\
 &=\sum_{m=1}^N \hat{\lambda}^{(l)}_m \overline{\beta^{(i)}_m}\beta^{(k)}_m.
  \end{split}
  \end{equation*}
In the above equations, $\lambda^{(l)}_m$ and $\hat{\lambda}^{(l)}_m$ are rotation terms. They are different for each $m$.  In both of the cases, the interference noise and the auto-correlation function are expressed as the weighted sum of the coefficients. 
  
\section{Conclusion}
In this paper, it has been shown that the Weyl spreading sequences appear as the orthogonal basis vectors in a chip-synchronous CDMA model. From this result, the interference noise always gets low when we use the Weyl spreading sequence. However, it shows that auto-correlation functions get high in all $l$ when we use the spreading sequence $\mathbf{w}_m(\sigma)$. The function $W_{i,i}(l)$ represents auto-correlation function. When we consider $B^{(l)}_{1,1}$ and $\mathbf{w}_{i}(0)$, the auto-correlation function is expressed as
\begin{eqnarray}
 W_{i,i}(l) &=& \mathbf{w}_i(0)^*  B^{(l)}_{1,1}\mathbf{w}_i(0)\nonumber\\
 &=&\mathbf{w}_i(0)^*  V \Lambda^{(l)} V^*\mathbf{w}_i(0)\nonumber\\
 &=& \mathbf{e}_i^{\mathrm{T}} \Lambda^{(l)} \mathbf{e}_i\nonumber\\
 &=&\lambda^{(l)}_i, \nonumber
 \end{eqnarray}
where $\mathbf{e}_i$ is the unit vector whose $i$-th element is $1$ and the other elements are $0$. The eigenvalues satisfy that $|\lambda^{(l)}_i| = 1$. Therefore, $|W_{i,i}(l)|$ becomes high in all $l$.

With the above result, it has been shown that any spreading sequence is expressed as the weighted sum of $\mathbf{w}_{m}(0)$ or $\mathbf{w}_{m}\left(1/(2N)\right)$. The interference noise $W_{i,k}(l)$ is expressed as the weighted sum of $\overline{\alpha^{(i)}_m}\alpha^{(k)}_m$ or $\overline{\beta^{(i)}_m}\beta^{(k)}_m$. This expression is useful for designing new spreading sequences since their coefficients naturally appear in the interference noise and the auto-correlation which are the key clue for designing. 
\references


\begin{thebibliography}{99}
\bibitem{gold}
R. Gold, Optimal binary sequences for spread spectrum multiplexing, {\it IEEE Transactions on Information Theory},
 13.4 (1967), 619-621.
\bibitem{chaos}
K. Umeno and K. Kitayama, Spreading sequences using periodic orbits of chaos for CDMA, {\it Electronics Letters} 35.7 (1999), 545-546.
 \bibitem{weylseq}
H. Tsuda and K. Umeno, Weyl Spreading Sequence Optimizing CDMA, arXiv preprint arXiv:1602.04584 (2016).
\bibitem{weyl}
H. Weyl, \"{U}ber die gleichverteilung von zahlen mod. eins, {\it Mathematische Annalen} 77.3 (1916), 313-352.
\bibitem{almost}K. Umeno, Spread Spectrum Communications Based on Almost Periodic Functions,  {\it IEICE Technical Report}, NLP 2014-101 (2014), 11-16.
\bibitem{pursley}M. B. Pursley,  Performance evaluation for phase-coded spread-spectrum multiple-access communication. I -system analysis, {\it IEEE Transactions on Communications} 25 (1977), 795-799.
\end{thebibliography}
\end{document}